\documentclass{pasj00}
\draft

\begin{document}
\SetRunningHead{Lee et al.}{Running Head}
\Received{--}%{yyyy/mm/dd}
\Accepted{--}%{yyyy/mm/dd}

\title{Nature of Infrared Sources in 11 $\mu$m Selected Sample from
 Early Data of the AKARI North Ecliptic Pole Deep Survey}

%%% begin:list of authors
\author{
    Hyung Mok \textsc{Lee}\altaffilmark{1},
    Myungshin \textsc{Im}\altaffilmark{1},
    Takehiko \textsc{Wada}\altaffilmark{2}
    Hyunjin \textsc{Shim}\altaffilmark{1},
    Seong Jin \textsc{Kim}\altaffilmark{1},
    Myung Gyoon \textsc{Lee}\altaffilmark{1},
    Narae \textsc{Hwang}\altaffilmark{1},
    Hideo \textsc{Matsuhara}\altaffilmark{2},
    Takao \textsc{Nakagawa}\altaffilmark{2},
    Shinki \textsc{Oyabu}\altaffilmark{2},
    Chris P. \textsc{Pearson}\altaffilmark{2},
    Toshinobu \textsc{Takagi}\altaffilmark{2},
    Takashi \textsc{Onaka}\altaffilmark{3},
    Naofumi \textsc{Fujishiro}\altaffilmark{2},
    Hitoshi \textsc{Hanami}\altaffilmark{4},
    Daisuke \textsc{Ishihara}\altaffilmark{3},
    Yoshifusa \textsc{Ita}\altaffilmark{2},
    Hirokazu \textsc{Kataza}\altaffilmark{2},
    Woojung \textsc{Kim}\altaffilmark{2},
    Toshio \textsc{Matusmoto}\altaffilmark{2},
    Hiroshi \textsc{Murakami}\altaffilmark{2},
    Youichi \textsc{Ohyama}\altaffilmark{2},
    Itsuki \textsc{Sakon}\altaffilmark{3},
    Toshiko \textsc{Tanabe}\altaffilmark{5},
    Kazunori \textsc{Uemizu}\altaffilmark{2},
    Munetaka \textsc{Ueno}\altaffilmark{6},
    Fumihiko \textsc{Usui}\altaffilmark{2}, and
    Hidenori \textsc{Watarai}\altaffilmark{7}
    }

\altaffiltext{1}{Astronomy Program, FPRD, Department of Physics and Astronomy,
\\ Seoul National University, Shillim-Dong, Kwanak-Gu, Seoul 151-742,
South Korea} \email{hmlee@snu.ac.kr; mim@astro.snu.ac.kr}
\altaffiltext{2}{Institute of Space and Astronautical
 Science, Japan Aerospace Exploration Agency,
\\ Yoshinodai 3-1-1, Sagamihara, Kanagawa 229-8510, Japan}
\altaffiltext{3}{Department of Astronomy, School of Science,
  University of Tokyo,
 \\ Bunkyo-ku, Tokyo 113-0033, Japan}
\altaffiltext{4}{Iwate University, 3-18-8 Ueda, Morioka 020-8550, Japan}
\altaffiltext{5}{Institute of Astronomy, University of Tokyo, 2-21-1
  Osawa, Mitaka, Tokyo, 181-0015, Japan}
\altaffiltext{6}{Department of Earth Science and Astronomy, University
  of Tokyo, 3-8-1, Komaba, Megro-ku, Tokyo 153-8902, Japan}
\altaffiltext{7}{Office of Space Applications, Japan Aerospace Exploration Agency,
\\  Tsukuba, Ibaraki, 305-8505, Japan}

%%% end:list of authors

%%% Please use the following style in case that sorting by
%%% affilation is impossible.
%
% \author{%
%   D-Firstname \textsc{D-Familyname}\altaffilmark{1}
%   E-Firstname \textsc{E-Familyname}\altaffilmark{1,2}
%   and
%   F-Firstname \textsc{F-Familyname}\altaffilmark{2}}
% \altaffiltext{1}{Address of Institute}
% \email{ddddd@xxx.xxx.xx.xx}
% \email{eeeee@xxx.xxx.xx.xx}
% \altaffiltext{2}{Address of Institute}
%% `\KeyWords{}' always has to be placed before `\maketitle'.
%Do NOT move this preamble from here!

\maketitle

\begin{abstract}
 We present the properties of 11 $\mu$m selected sources detected
in the early data of the North Ecliptic Pole Deep (NEP-Deep) Survey of
AKARI. The data set covers 6 wavelength bands from 2.5 to 11
$\mu$m, with the exposure time of 10 $\sim$ 20 minutes. This field
lies within the CFHT survey with four filter bands ($g^\prime,
r^\prime, i^\prime ,z^\prime$), enabling us to establish nearly
continuous spectral energy distributions (SEDs) for wavelengths
 ranging from 0.4 to 11 $\mu$m.
  The main sample studied here
 consists of 72 sources whose 11 $\mu$m AB magnitudes
 are equal to or brighter than 18.5 (251 $\mu$Jy), which is
 complete to more than 90\%.
  The 11 $\mu$m band has an advantage of sampling star forming
 galaxies with low to medium redshifts since the prominent PAH
 feature shifts into this band.
 As expected, we find that the majority ($~69\%$) of 11 $\mu$m bright
 sources are star forming galaxies at $0.2 \lesssim z \lesssim 0.7$
 with $L_{IR} \sim 10^{10}$
 -- $10^{12}~L_{\odot}$ based on the detailed modelling of SEDs.
 We also find four AGNs lying at various redshifts in the
 main sample. In addition,
 we discuss a few sources which have non-typical SEDs of the main
 sample, including a brown dwarf candidate, a steep power-law source,
 flat spectrum object, and an early-type galaxy at moderate redshift.
\end{abstract}

\section{INTRODUCTION}\label{sec:introduction}

  AKARI is an infrared space telescope, which was launched in
 February 2006, in order to carry out all sky survey at mid- to
 far-infrared (\cite{mura98}; Nakagawa 2001, \cite{shib04}). In
addition to the all sky survey, AKARI is capable of making pointed
observations at near- to far-infrared. Because of the nature of
the sun-synchronous polar orbit, North and South Ecliptic Polar
regions (NEP and SEP, hereafter) have very good visibility so that
repeated exposures are possible. In particular, an extragalactic
 survey of the NEP region has been designed to perform the deep
 imaging of pre-selected blank field from 2.5 to 26 $\mu$m
in order to study the formation and evolution of galaxies. More
details of the deep extragalactic NEP survey (hereafter, NEP-Deep)
 and its strategies can be found in
Matsuhara et al. (2006).
 Currently, we are planning to release the NEP survey data to public
 in 2008, approximately one year after the completion of the survey.

 Prior to the full, complete NEP survey, mini surveys have been
carried out in order to assess the capabilities of AKARI.
  One of them is an imaging survey
 called the NEP-deep-early survey. The NEP-deep-early field covers
 two fields of views,  one with the MIR-L camera (L15, L18W, and L24 bands),
 and another with the NIR and the MIR-S
 cameras with 6 bands covering 2.5 to 11
 $\mu$m,
 at the total accumulated on-source exposure time of 10 to 20 minutes using
IRC05 Astronomical Observation Template (AOT) (Onaka et al. 2007).
 The size of each field is roughly 10$^\prime \times 10^\prime$.

  The NIR/MIR-S part
 is covered by the CFHT Megaprime optical
 survey at $g^\prime$, $r^\prime$, $i^\prime$, and $z^\prime$ bands
 (Hwang et al. 2007).
  Other mini surveys include the NEP monitor field
 survey taken with a different observing mode,
 and the AKARI/NEP performance verification (NEP-PV) field survey.

 The purpose of the current paper is to
 examine the properties of the infrared sources detected in the
 NIR/MIR-S part of the NEP-deep-early survey field.
  While other survey fields provide a wider wavelength coverage (the
 NEP monitor field survey)
 or different wavelength information (L15; NEP-PV survey),
  the NEP-deep-early field data can provide with
 the complementary information regarding the nature of
 infrared sources expected in the NEP wide survey
 which has a similar survey depth to
 this data set but with much wider areal coverage.
 Analysis of an independent
 field also provides an opportunity to discover interesting rare sources.
  In particular, we will study in this paper, the nature of
 11 $\mu$m selected objects. Related topics will be covered
 by other papers: detailed analysis of the monitor field data
 (Takagi et al. 2007), the optical identification of 15 $\mu$m
 selected sources (Matsuhara et al. 2007), the source
 counts in N3, S7, and L15-bands (Wada et al. 2007), and
 X-ray sources in the NEP survey data (Oyabu et al. 2007).

\begin{figure}
 \begin{center}
    \FigureFile(130mm, 150mm){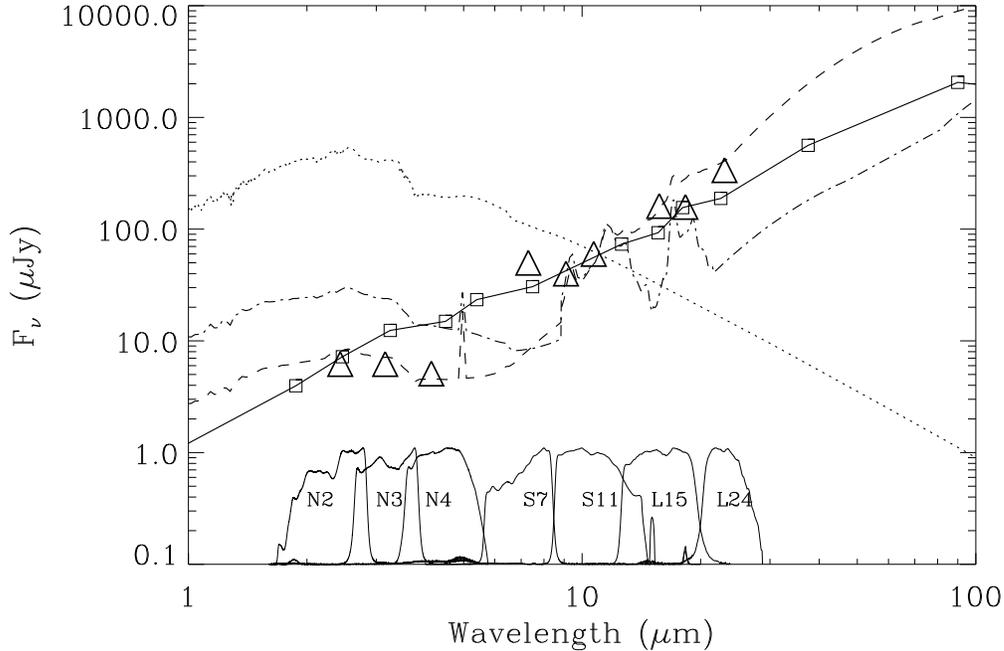}
 \end{center}
 \caption{Spectral energy distribution of various sources normalized
 to the AKARI IRC detection limits of our data at 11 $\mu$m.
  Different lines represent SEDs of different types of galaxies
 at $z=0.5$ which is a typical redshift of 11 $\mu$m sources:
 a passively evolving galaxy
 with the age of 2.5 Gyr (dotted line), a galaxy with moderate star
 formation rate like typical disk galaxies (dot-dashed line),
 infrared luminous starburst galaxy similar to Arp220 (dashed line),
 and an AGN, Mrk231 (small squares connected with solid lines).
  Also plotted are the detection limits (triangles), and the
 filter response curves. This figure demonstrates that S11 is effective
 at picking up the PAH feature of a moderate redshift star-forming galaxy.
  Note that S9W covers the spectral responses of S7 and S11.}
\end{figure}

  Among the six wavelength bands of the NEP-deep-early survey data, the 11
 $\mu$m image has been used in the primary selection of the sample.
   The S11 passband samples the wavelength red-ward of the
 Spitzer IRAC channel 4 (8 $\mu$m) from 8 $\mu$m to 15 $\mu$m,
 filling in a gap in the Spitzer passbands.
  At 11 $\mu$m, there is a strong silicate absorption in
 star forming galaxies at $z=0$, but
 the S11 band is also where the strong PAH features at 6.3 and 7.7 $\mu$m
 redshift into at $0.2 \lesssim z \lesssim 0.7$ (Fig. 1).
  Therefore, unlike source selections based on an 8 $\mu$m
 band (e.g., the Spitzer IRAC Channel 4), the 11 $\mu$m sources
 are likely to be dominated by the star-forming galaxies at redshift
 in the range 0.2 $\sim$ 0.7.
   Previous studies of IR luminous galaxies and star formation
 activities using the {\it Spitzer} data have relied upon the extrapolation
 of the 8 micron and 24 micron
 bands, which can introduce a significant
  uncertainty in the estimate of IR luminosity. With the unique
 capability of the 11 $\mu$m imaging of AKARI, we should be
 able to reduce such an uncertainty in understanding the IR activity
 of galaxies at moderate redshifts.
  Other sources that can be found in the 11 $\mu$m include
 Active Galactic Nuclei (AGNs) and stars.

  In this work, we characterize properties of objects selected based upon
 this new wavelength regime.
 This paper is organized as follows. In the immediately following
section, we describe the details of the NEP-deep-early survey data set.
 In \S 3, we present the results of the photometry on the data. The
characteristics of the infrared sources in conjunction with other
available data are discussed in detail in \S 4. In addition to the
11 $\mu$ flux limited sample, we also find sources that need
special attention based on the SEDs in other bands. Such objects
are discussed in \S5. The final section summarizes our main
results.
  Throughout this paper, we will use AB magnitude system which
 converts to $\mu$Jy as ABmag=$23.9 - 2.5 \log_{10}[Flux (\mu {\rm Jy)}]$.

\section{NEP-DEEP-EARLY SURVEY DATA}\label{sec:data}

 The NEP-deep-early survey data have been taken during the early phase
of the scientific observations that started in early May 2006
in order to assess the performance of the IRC
instrument (Onaka et al. 2007).
 The passbands covered by this mini survey are N2,
N3, N4, S7, S9W, and S11 (see \cite{matsu06} for the detailed
information on the designation of the filters). Our field
 is centered on $\alpha = 17^h 56^m 48^s$, $\delta = + 66^\circ 09^\prime
 49^{\prime\prime}$ with approximately $10^\prime \times 10^\prime$ field
 of view.
This area lies within the proposed NEP-Deep survey of 0.5 deg$^2$
circular field centered on $\alpha= 17^h 55^m 24^s$, $\delta =
+66^\circ 37^\prime 32^{\prime\prime}$.

\begin{table}
\caption{Observational Parameters of NEP Early Data}
\label{table1}
\begin{center}
\begin{tabular}{c c c c}
\hline Filter Name & Effective Wavelength ($\mu$m) & Total
Integration Time (sec)&
Detection Limit ($5 \sigma$, $\mu$Jy: Estimates) \\
\hline \hline
N2 &2.43 & 654.5 & 6.2\\
N3 &3.16 &981.7 & 6.2\\
N4 &4.14 &589.1 & 5.1 \\
S7 &7.19 & 1227.0& 50 \\
S9W &8.74 &768.5 & 40 \\
S11 &10.4 &687.1 & 60 \\
\hline

\end{tabular}
\end{center}
\end{table}

  The data were taken with the IRC05 AOT mode which uses a rather
 small number of resets per pointing opportunity that lasts for
 about 10 minutes in order to achieve deep imaging. This AOT also
 uses Fowler 16 data reading scheme which increases the
 signal-to-noise ratio significantly. The effective on-source
 exposure time of 1 pointing observation is a bit over 5 minutes.
  The observing parameter of each pointing is listed in Table 1.

  The frames taken in a single observing session that lasts for about
 10 minutes are processed and stacked by using
 the IRC data reduction pipeline version 070104 (Ita et al. 2007).
  Within this pipeline, the anomalous pixels have been clipped,
 dark signals have been subtracted with the darks taken in each pointing
 observation (self-dark), and the linearity correction has been
 carried out.
  The optical distortion due to off-axis imaging
 has also been corrected for. The cosmic rays have been removed at
 this stage.

  The images taken during different pointing observations
 have been co-added by using SWarp\footnote
 {http://terapix.iap.fr/rubrique.php?id\_rubrique=49}.
 The result of the coadded master image at S11 band is shown in
  Fig. 2, together with the g-r-i color composite of the optical image
 taken with the CFHT Megaprime (Hwang et al. 2007).
  The depth of the optical data is 26.1, 25.6, 24.9, and 23.7
 AB magnitude at 4-$\sigma$ over an 1\farcs0 aperture in
 $g^{'}, r^{'}, i^{'}$, and $z^{'}$ bands.
   The optical data allow us to provide further morphological information
 on the sources detected in the infrared.
  We use the CFHT data set for identification of infrared sources as
 well as to construct continuous SEDs of detected sources from
 optical to mid-infrared wavelengths.

\begin{figure}
\begin{center}
 \FigureFile(105mm, 120mm){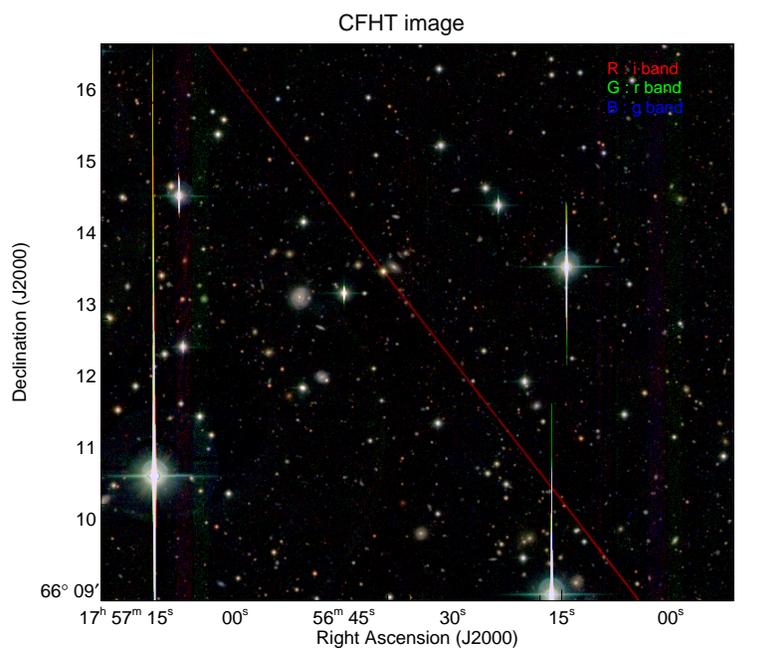}
 \FigureFile(105mm, 120mm){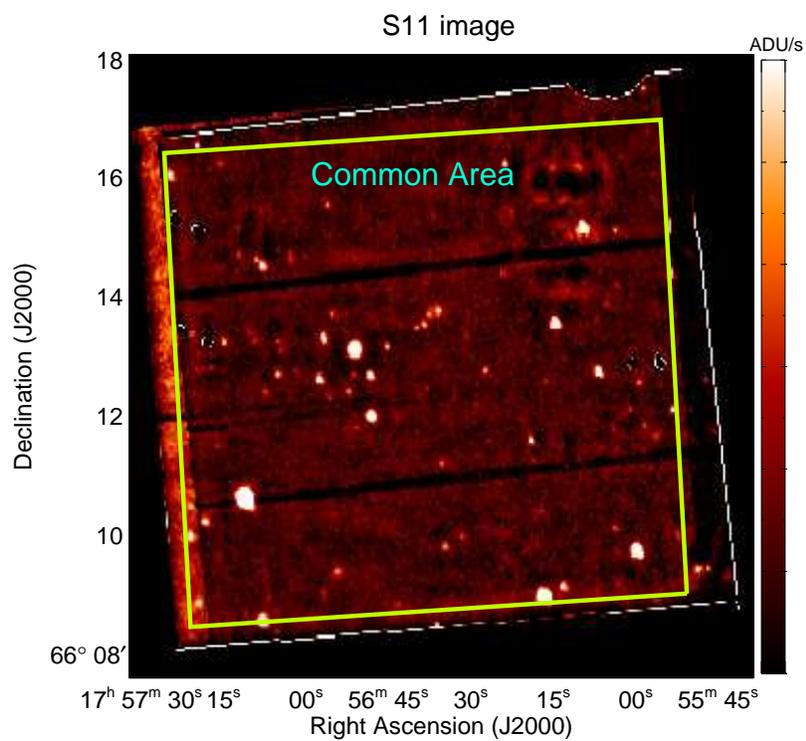}
\end{center}
\caption{The RGB representation of CFHT image (upper panel) and
the master image which was produced by combining all exposures
with S11 filter. The total exposure time for S11 image image is
about 10 minutes.}
\end{figure}

\section{Detection and Selection of Sources}\label{sec:detection}

  Source detection was done by using SExtractor
 (Bertin \& Arnouts 1996) for the images stacked with
 SWarp, using the following parameters for the detection:
 {\tt DETECT\_MINAREA=5., DETECT\_THRESH=1.5, BACKSIZE=16,
 BACK\_FILTERSIZE=3.} These parameters correspond roughly to
 the S/N cut of 3.4.
  From the background noise, we estimate that the 5-$\sigma$ detection
 limit of sources over 7\farcs0 diameter aperture to be 60 $\mu$Jy
 or 19.5 AB magnitude.

  The conversion to Jy unit and AB magnitude from the
 measured DN was done by applying the flux conversion factor
 provided in the IRC manual version 1.1, released on 2007 March 22.
  When converting DN to flux, we also applied the color correction
 factor provide in the IRC manual since the color correction can
 change the flux value by as much as 30\% in some cases. The application
 of the color correction is not straightforward because of the
 complex shape of the spectral energy distribution of 11 $\mu$m
 sources. The IRC manual provide the color correction for black body
 with various temperatures. As a first-order approximation,
 we adopted the color correction factor of a black body with
 the temperature which matches the observed color of the object at the
 wavelengths of the passband in question and the passband
 closest to it. For example, the color correction factor of
 S7-band is derived by matching the observed S7-S9W color with
 the same color of a black body. For N3 and S7, we adopted the
 mean of the color correction factors derived from N2-N3/N3-N4 and
 S7-S9W/S9W-S11 respectively. With the color correction factor,
 we estimate the accuracy of the flux calibration to be
 roughly 15 \% from  stellar spectra (see Section 4.4).

  Not all of the detected sources are
 real objects, and our test of the reliability of source detection on
 the negative image -- the S11 image that has been multiplied by -1 --
 shows that there are spurious sources even at a very bright
 magnitude due to several instrumental effects such as
 imperfect flat image. Therefore, we excluded spurious
 sources manually by checking the S11 detection against the AKARI NIR
 images and CFHT optical images for objects brighter than 19 AB mag objects.
 If an S11 detection accompanies detection in none of the AKARI NIR
 and the CFHT bands,
 we considered the source detection to be spurious. Detection in the shorter
 MIR-S bands alone is not considered to be true, since they can be ghost images
 from the shared MIR-S optics.
  After the manual inspection, the final list of S11 sources includes 176,
 among which 113 are brighter than 19.0 mag.

  We use the AUTO-magnitude from the SExtractor for the S11 flux.
  We have compared the auto-magnitude against aperture magnitude
 with $\sim$7.2 arcsec diameter and an aperture correction derived from
 the growth curve, and found a reasonable agreement between the two.
  When there are close neighbors that could affect the AUTO-magnitude
 (i.e., the source being likely to be blended one), we used the aperture
 magnitude with the aperture diameter of 7\farcs2.
  Based on the 11 $\mu$m-selected sample, we have produced a
 band-merged catalogue as follows. In order to measure fluxes
 in the other IRC bands, we used
 ASSOC\_parameters in SExtractor. Fluxes were measured at the
 position of each object over the same area used for
 the S11 flux measurement, after finding the nearest counterpart
 within 4\farcs0 radius.
  In case of the CFHT optical bands, 11 $\mu$m source
 catalog was matched with the CFHT catalog with a matching
 radius of 3\farcs0. The matching shows that almost all of
 the matched CFHT sources lie within 2\farcs0 from the S11 position.
  When multiple objects are matched for a single S11 source,
 images were examined by eyes to find the most probable object that is
 responsible for the 11 $\mu$m detection.
  Since the PSF shape varies among different AKARI passbands,
 our procedure may introduce additional error in the flux measurement
 at passbands below 11 $\mu$m. We have checked whether such an error
 is significant by performing the SExtractor detection and flux estimate
 on NIR images, and comparing the resultant AUTO-flux with the flux
 measured through the above procedure. We find that
 the additional uncertainty introduced by the above flux estimate to be less
 than 10\%.

  We have divided the detected sources at 11 $\mu$m into stellar
 and non-stellar sources, based on the cross identification with
 the CFHT data.
  Through visual inspection, we classified an object as a star when the object
 has a clear trace of point sources such
 as diffraction spikes and has the minimal FWHM width. The stellarity
 values are checked for the sources that appear to be a point source.

\begin{table}
%\caption{List of Sources brighter than $S11=18.5^m$}

\label{table2}
\begin{center}
Table 2. List of Sources brighter than $S11=18.5^m$
\vskip 0.2cm
\begin{footnotesize}
\begin{tabular}{c c c c c c c c c l}
\hline \hline
Id & RA & Dec & N2 & N3 & N4 & S7 & S9W & S11 & Type \\
(1) & (2) & (3) & (4) & (5) & (6) & (7) & (8) & (9) & (10) \\
\hline
J175545.68+661144.7        &   268.94034  &   66.1958     &   19.93   &   19.89   &   19.94   &   20.5   &   19.8   &   18.1   &   SF Galaxy \\
J175546.49+660926.2        &   268.94373  &   66.1573     &   18.38   &   18.47   &   18.90   &   17.9   &   18.0   &   17.3   &   SF Galaxy \\
J175546.64+661236.8        &   268.94434  &   66.2102     &   18.90   &   19.11   &   20.10   &   19.4   &   19.1   &   17.9   &   SF Galaxy \\
J175548.42+660925.6        &   268.95175  &   66.1571     &   18.78   &   18.86   &   19.06   &   19.0   &   *   &   18.0   &   SF Galaxy \\
J175552.03+661000.7        &   268.96680  &   66.1669     &   19.56   &   19.82   &   20.11   &   18.1   &   17.0   &   16.5   &   SF Galaxy \\
J175552.67+661421.6        &   268.96948  &   66.2393     &   17.55   &   17.80   &   18.23   &   18.3   &   17.6   &   17.0   &   SF Galaxy \\
J175552.76+661205.8        &   268.96985  &   66.2016     &   18.53   &   18.86   &   19.17   &   18.7   &   17.6   &   17.2   &   SF Galaxy \\
J175556.06+661139.1        &   268.98361  &   66.1942     &   19.83   &   19.78   &   20.11   &   *   &   19.0   &   17.8   &   SF Galaxy \\
J175556.11+661343.1        &   268.98380  &   66.2286     &   19.42   &   19.60   &   19.76   &   19.4   &   18.1   &   18.1   &   SF Galaxy \\
J175556.59+661539.9        &   268.98581  &   66.2611     &   20.92   &   21.16   &   20.99   &   *   &   19.7   &   18.4   &   SF Galaxy \\
J175558.04+661150.8        &   268.99185  &   66.1975     &   17.58   &   17.81   &   18.46   &   19.5   &   18.8   &   18.0   &   SF Galaxy \\
J175559.06+661235.0        &   268.99612  &   66.2097     &   18.77   &   18.89   &   19.17   &   20.1   &   19.2   &   17.8   &   SF Galaxy \\
J175559.52+660944.2        &   268.99802  &   66.1623     &   17.02   &   17.25   &   17.45   &   *   &   16.1   &   14.8   &   SF Galaxy \\
J175602.13+661620.2        &   269.00888  &   66.2723     &   19.78   &   19.84   &   20.59   &   *   &   19.8   &   18.3   &   SF Galaxy \\
J175603.06+660912.4        &   269.01279  &   66.1534     &   19.45   &   19.70   &   20.15   &   20.4   &   18.4   &   18.4   &   SF Galaxy \\
J175603.45+661506.5        &   269.01440  &   66.2518     &   19.16   &   18.77   &   19.03   &   18.9   &   18.7   &   18.2   &   SF Galaxy \\
J175603.58+661007.7        &   269.01492  &   66.1688     &   19.01   &   19.15   &   19.47   &   18.2   &   17.6   &   18.5   &   SF Galaxy    \\
J175606.54+661129.6        &   269.02728  &   66.1916     &   17.14   &   17.48   &   18.43   &   18.7   &   19.3   &   18.3   &   Star \\
J175606.58+661244.6        &   269.02744  &   66.2124     &   18.66   &   17.85   &   17.54   &   16.6   &   16.5   &   16.2   &   AGN \\
J175609.47+661508.9        &   269.03949  &   66.2525     &   17.38   &   16.69   &   16.58   &   15.9   &   15.6   &   15.4   &   AGN \\
J175613.04+660907.8        &   269.05438  &   66.1522     &   16.34   &   16.48   &   17.19   &   17.6   &   17.4   &   16.7   &   SF Galaxy    \\
J175614.60+661332.0        &   269.06085  &   66.2256     &   12.79   &   13.18   &   13.83   &   14.7   &   15.2   &   15.6   &   Star \\
J175616.01+661235.3        &   269.06674  &   66.2098     &   20.20   &   19.68   &   19.60   &   19.5   &   19.7   &   18.2   &   SF Galaxy  \\
J175617.63+660951.6        &   269.07349  &   66.1643     &   18.77   &   18.11   &   19.82   &   18.6   &   *   &   18.2   &   Early-type \\
J175618.63+661136.6        &   269.07764  &   66.1935     &   18.31   &   18.61   &   19.00   &   18.2   &   17.4   &   17.4   &   SF Galaxy \\
J175622.40+661525.4        &   269.09338  &   66.2570     &   19.10   &   19.19   &   19.46   &   19.0   &   17.8   &   17.6   &   SF Galaxy \\
J175623.97+661423.9        &   269.09988  &   66.2400     &   15.38   &   15.87   &   16.57   &   17.7   &   18.7   &   18.3   &   Star \\
J175623.98+661613.6        &   269.09992  &   66.2705     &   18.07   &   18.34   &   18.78   &   19.6   &   17.5   &   17.8   &   SF Galaxy \\
J175624.41+661641.5        &   269.10172  &   66.2782     &   19.73   &   19.40   &   19.49   &   19.7   &   19.5   &   18.3   &   SF Galaxy  \\
J175627.10+661245.3        &   269.11295  &   66.2126     &   21.89   &   21.75   &   20.22   &   18.6   &   18.4   &   17.8   &   Early-type  \\
J175631.06+661653.8        &   269.12946  &   66.2816     &   16.57   &   17.00   &   17.54   &   18.6   &   19.0   &   18.4   &   Early-type \\
J175631.47+660958.9        &   269.13117  &   66.1664     &   19.78   &   19.75   &   20.01   &   20.4   &   *   &   18.4   &   SF Galaxy   \\
J175634.29+660846.9        &   269.14288  &   66.1464     &   15.69   &   16.26   &   16.95   &   *   &   16.9   &   17.7   &   Star \\
J175634.44+660949.3        &   269.14352  &   66.1637     &   16.84   &   17.11   &   17.81   &   18.8   &   17.6   &   16.5   &   SF Galaxy \\
J175635.13+661112.9        &   269.14639  &   66.1869     &   20.02   &   19.90   &   20.50   &   20.8   &   19.7   &   18.0   &   SF Galaxy    \\
J175635.94+660824.2        &   269.14975  &   66.1401     &   13.58   &   13.36   &   13.77   &   14.4   &   15.7   &   16.3   &   Star  \\
J175637.53+661342.9        &   269.15640  &   66.2286     &   17.01   &   16.98   &   17.44   &   17.2   &   17.6   &   18.1   &   early-type \\
J175639.09+661547.2        &   269.16290  &   66.2631     &   17.86   &   18.40   &   18.83   &   19.6   &   *   &   17.7   &   SF Galaxy    \\
J175639.80+661328.8        &   269.16583  &  66.2247    & 15.31     &   15.73   &  16.40   &  17.4 &   17.9 &   18.5  &  Star \\
J175643.10+660850.4        &   269.17960  &   66.1473     &   19.46   &   19.83   &   19.90   &   *   &   18.2   &   17.5   &   SF Galaxy    \\
J175645.21+661310.5        &   269.18839  &   66.2196     &   14.70   &   15.16   &   15.85   &   16.9   &   17.2   &   18.0   &   Star \\
J175648.21+661200.1        &   269.20090  &   66.2000     &   16.61   &   17.04   &   17.55   &   17.1   &   16.4   &   16.2   &   SF Galaxy \\
\hline
\end{tabular}
\end{footnotesize}
\end{center}
\end{table}
\break

\begin{table}
\begin{center}
\centerline{ Table 2 -- continued}
\vskip 0.2cm
\begin{footnotesize}
\begin{tabular}{c c c c c c c c c l}
\hline
\hline Id & RA & Dec & N2 & N3 & N4 & S7 & S9W & S11 & Type \\
(1) & (2) & (3) & (4) & (5) & (6) & (7) & (8) & (9) & (10) \\
\hline

J175648.43+661241.7        &   269.201843  &   66.2116     &   18.12   &   18.48   &   18.54   &   18.0   &   17.1   &   16.8   &   SF Galaxy \\
J175651.33+661307.2        &   269.213898  &   66.2187     &   15.43   &   15.83   &   16.24   &   16.2   &   15.3   &   15.1   &   SF Galaxy \\
J175651.45+661242.5        &   269.214386  &   66.2118     &   20.06   &   20.26   &   19.58   &   19.2   &   19.4   &   18.4   &   AGN \\
J175652.34+661225.3        &   269.218109  &   66.2070     &   20.38   &   19.96   &   20.31   &   19.5   &   18.6   &   17.9   &   SF Galaxy \\
J175654.21+660924.1        &   269.225891  &   66.1567     &   18.92   &   18.89   &   19.48   &   18.2   &   17.6   &   17.2   &   SF Galaxy \\
J175655.07+661343.7        &   269.229462  &   66.2288     &   18.09   &   18.27   &   18.90   &   19.0   &   18.6   &   18.0   &   SF Galaxy \\
J175655.97+661534.1        &   269.233246  &   66.2595     &   20.39   &   19.71   &   19.29   &   18.7   &   17.9   &   17.9   &   AGN \\
J175656.50+661316.0        &   269.235413  &   66.2211     &   19.06   &   18.95   &   19.11   &   17.5   &   17.3   &   16.6   &   SF Galaxy \\
J175657.46+661237.6        &   269.239441  &   66.2105     &   18.08   &   18.37   &   18.73   &   18.6   &   17.0   &   16.6   &   SF Galaxy \\
J175658.82+660834.6        &   269.245117  &   66.1429     &   18.31   &   18.45   &   18.88   &   18.5   &   18.2   &   17.9   &   SF Galaxy \\
J175702.81+661439.5        &   269.261749  &   66.2443     &   19.36   &   19.22   &   19.43   &   19.7   &   18.7   &   18.0   &   SF Galaxy \\
J175704.53+661304.9        &   269.268890  &   66.2180     &   18.13   &   18.36   &   18.95   &   19.7   &   18.9   &   18.4   &   SF Galaxy \\
J175705.12+661242.2        &   269.271362  &   66.2117     &   18.72   &   18.44   &   18.64   &   18.4   &   18.3   &   17.6   &   SF Galaxy   \\
J175705.16+661316.2        &   269.271515  &   66.2212     &   18.37   &   18.60   &   19.20   &   20.0   &   *   &   18.1   &   SF Galaxy \\
J175707.37+661224.7        &   269.280731  &   66.2069     &   15.51   &   15.96   &   16.65   &   17.8   &   18.0   &   17.6   &   Star \\
J175707.81+660833.9        &   269.282562  &   66.1427     &   16.08   &   16.60   &   17.13   &   *   &   *   &   16.1   &   SF Galaxy   \\
J175708.01+660928.6        &   269.283386  &   66.1579     &   19.72   &   19.20   &   19.40   &   18.4   &   17.5   &   18.2   &   SF Galaxy    \\
J175708.04+661431.5        &   269.283508  &   66.2421     &   13.96   &   14.45   &   15.08   &   15.8   &   16.5   &   17.0   &   Star \\
J175709.08+661439.2        &   269.287872  &   66.2442     &   15.21   &   15.73   &   16.46   &   *   &   *   &   17.8   &   Star \\
J175710.35+661324.3        &   269.293152  &   66.2234     &   16.34   &   16.61   &   16.83   &   18.0   &   18.0   &   17.5   &   SF Galaxy \\
J175712.35+661442.9        &   269.301483  &   66.2453     &   18.74   &   19.04   &   19.44   &   19.5   &   18.6   &   18.4   &   SF Galaxy    \\
J175715.07+661315.0        &   269.312805  &   66.2208     &   18.41   &   19.07   &   19.24   &   18.4   &   18.8   &   17.4   &   SF Galaxy \\
J175715.88+661614.8        &   269.316193  &   66.2708     &   19.58   &   19.29   &   19.68   &   20.9   &   18.8   &   17.4   &   SF Galaxy  \\
J175716.24+661131.3        &   269.317688  &   66.1920     &   15.58   &   15.97   &   16.65   &   17.8   &   17.7   &   18.2   &   Star    \\
J175718.52+661014.4        &   269.327210  &   66.1707     &   16.13   &   *   &   17.14   &   17.3   &   17.5   &   16.9   &   SF Galaxy \\
J175719.70+660852.0        &   269.332092  &   66.1478     &   18.16   &   *   &   18.27   &   18.7   &   *   &   17.4   &   SF Galaxy   \\
%J175720.16+660846.93        &   269.334015  &   66.1464     &   18.45   &   *   &   18.91   &   *   &   *   &   *   &   SF Galaxy \\
J175720.25+661630.9        &   269.334412  &   66.2753     &   14.45   &   14.97   &   15.65   &   16.7   &   17.5   &   17.9   &   Star \\
%J175720.84+661324.41        &   269.336884  &   66.2234     &   19.33   &   18.81   &   19.12   &   19.69   &   18.11   &   18.65   &   SF Galaxy \\
J175721.28+660959.0        &   269.338684  &   66.1664     &   14.42   &   *   &   15.90   &   16.4   &   17.2   &   17.4   &   Star    \\
J175725.25+661602.0        &   269.355225  &   66.2672     &   16.88   &   *   &   17.85   &   18.6   &   17.7   &   16.9   &   SF Galaxy \\
J175728.03+661530.0       &   269.366791  &   66.2583     &   20.47   &   *   &   *   &   *   &   *   &   18.1   &   SF Galaxy \\

\hline
\end{tabular}
\end{footnotesize}
\end{center}
{\footnotesize (1) ID of the object; (2) RA and DEC in J2000; (3)-(9):
AB magnitudes at these bands;
(10) Object type: SF Galaxy for star forming galaxy (Section 4.3.1),
 Early-type for a galaxy with E/S0-like spectra (Section 5),
 Star for point source with stellar spectra
 (Section 4.3.3), and AGN
 for power-law sources which are likely to be AGNs (Section 4.3.2).}
\end{table}

\section{RESULTS}

\subsection{S11 Number Count and the Magnitude Limited Sample}

  The number count of S11 sources as a function of AB magnitude
 is shown in Fig. 3. The total count (black histogram)
 as well as the count for non-stellar objects only
 (red histogram) are shown together. The expected distribution of
 sources uniformly distributed over the sky in a non-expanding,
 Euclidean universe is indicated as a straight line with slope of 0.6.
  The observed source count peaks at around 19-th magnitude
 and declines afterward. Clearly, we encounter the incompleteness
 effect beyond ~19-th magnitude. In order to investigate
 the incompleteness in our sample, we created artificial objects
 that have a Gaussian PSF shape with the FWHM value of the
 point sources in the S11 image. We distributed them
 in less crowded parts of the S11 image and performed the source detection.
 From this exercise, we find that our source counts is nearly
 complete down to $S11 \sim 18.5$ mag ($> 90$\%), while it drops to
 50\% level at 19.2 mag.

  Our visual inspection of S11 image also confirms that
 the number of spurious sources increases as we go to $S11 > 18.5$ mag.
  Therefore, we consider a sample with $S11 < 18.5$ mag to be
 highly complete and reliable, and use this sample of
 72 objects to characterize
 the property of S11 sources.
 The AB magnitudes of our sample are listed in Table 2,
 together with our identifications regarding the type of the objects from
 the SED shape as well as the appearance of the object. We have
 classified our sources into 4 categories: stars, star forming galaxies,
 early type galaxies, and AGNs.
  We also add a supplemental sample
 made of sources fainter than $S11 = 18.5$ mag, but have
 spectral energy distributions worth of special note. These
 sources are treated separately in \S5. We concentrate on the flux
 limited sample at 11 $\mu$m in this section.

\begin{figure}
  \begin{center}
    \FigureFile(130mm, 150mm){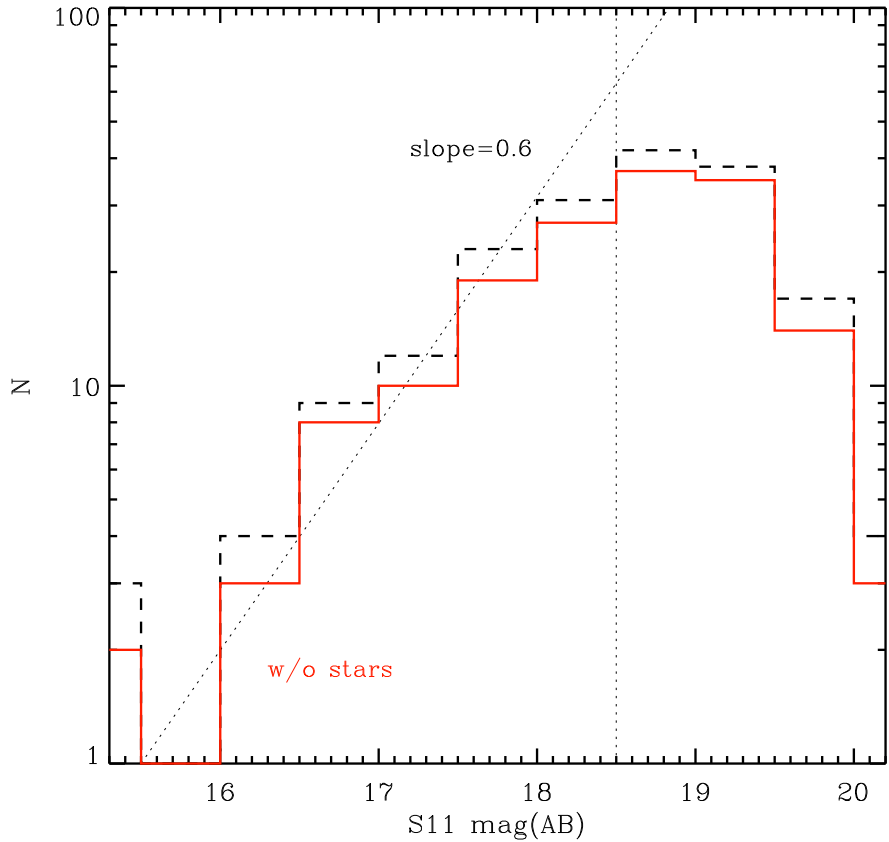}
  \end{center}
   \caption{The number distribution of S11 sources as a function of magnitude. Both
   total sources as well as non-stellar sources are plotted. At bright end, the number
   count approximately follows the $N \propto m^{0.6}$ power-law, but begins to deviate as the
   incompleteness becomes important at around 19-th magnitude. The stellar sources
   comprise only small fraction of the entire sample.}
   \label{counts}
\end{figure}

\subsection{Colors}

 The S7-S11 versus N2-N4 color-color diagram of the 11 $\mu$m flux
limited sample is shown in Fig. 4. The sources identified as stars
are indicated as asterisks while non-stellar sources are shown as
filled dots with different sizes depending on the brightness. Also
shown in this diagram is the same color-color sequences of three
typical nearby star forming galaxies M51, M82 and Arp 220 located
at redshifts in the range between $z=0$ and 2.2.
It is clear from this figure that most of the  S11 flux limited sources
 found in a shallow NEP survey are likely to be star
forming galaxies at low to medium redshifts. However, there are
also redder sources in N2-N4 color index than star forming spiral
galaxies. As we shall see, most of them are AGN.

\begin{figure}
  \begin{center}
    \FigureFile(130mm, 150mm){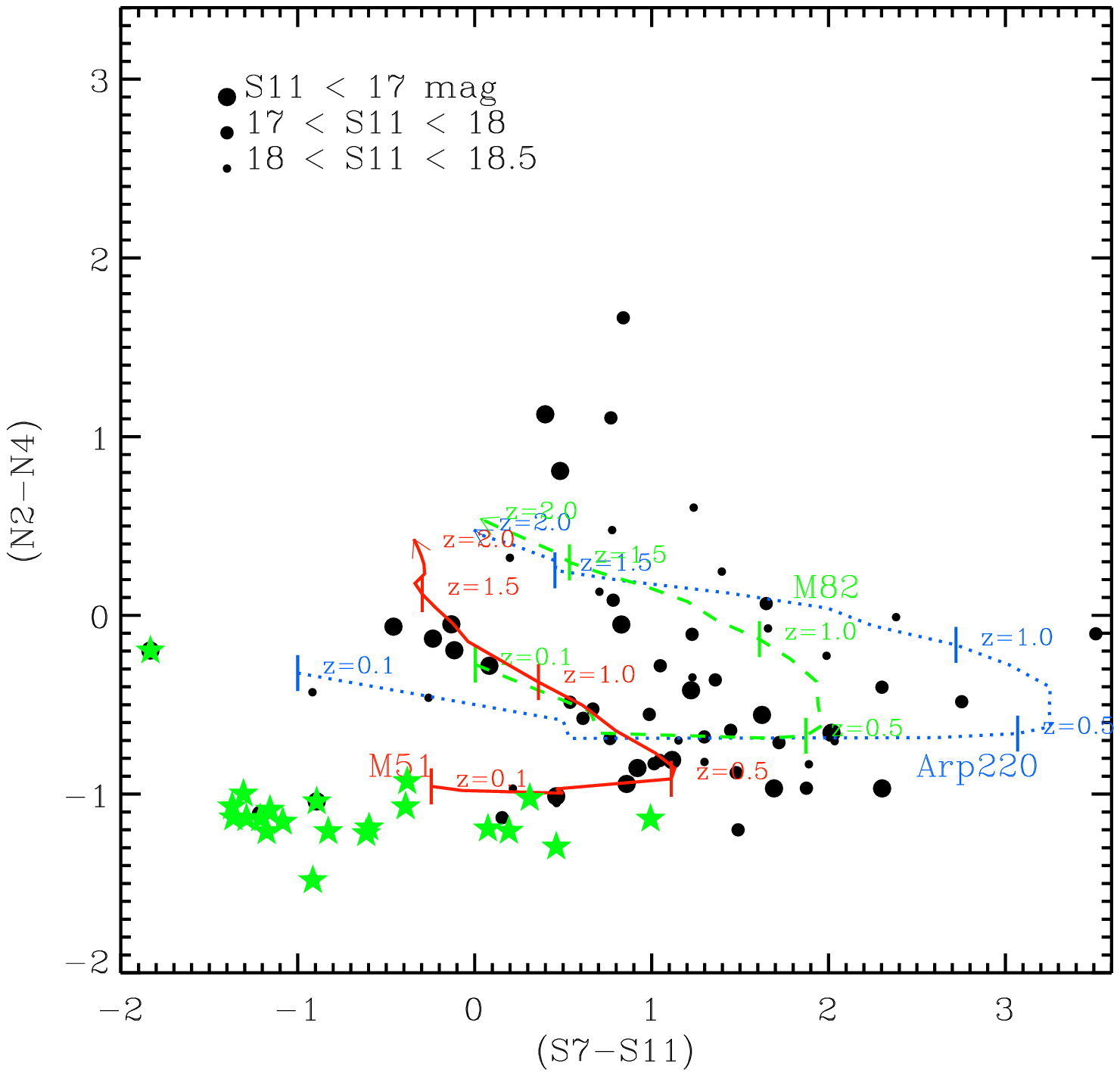}
  \end{center}
   \caption{The S7-S11 versus N2-N4 color-color plots of the S11 flux limited
 sample. The stars are located in the lower-left corner (green star symbols).
 Also shown in this diagram are the sequence of three star forming
 galaxies M51, M82 and Arp 220 at redshifts from z=0 to 2.2.
  Majority of the non-stellar sources are likely to be galaxies at
 low to moderate redshifts. AGNs occupy the upper right part of the
 color-color diagram because of their power-law SEDs.
}
   \label{cc_diagram}
\end{figure}

 The N4 versus N4-N2 and N4 versus N4-N3 color-magnitude diagrams
are shown in Fig. 5. In near infrared, the stellar SEDs are nearly
that of Rayleigh Jeans tail of black body, and therefore stars lie
in rather narrow range in this diagram. Also shown as solid and
dashed lines are locations of an $L^{*}$ galaxy having the SED
shape of M51 and M82, respectively, at different redshifts as
marked on the lines. Actual data points are consistent with the
star forming disk galaxies lying within the redshift $\lesssim 1$.
Again, the objects with very red color in N2-N4 are likely to be
AGNs. Such red objects are more easily distinguished in N2-N4 than
N4-N3.
 \begin{figure}
  \begin{center}
    \FigureFile(100mm, 100mm){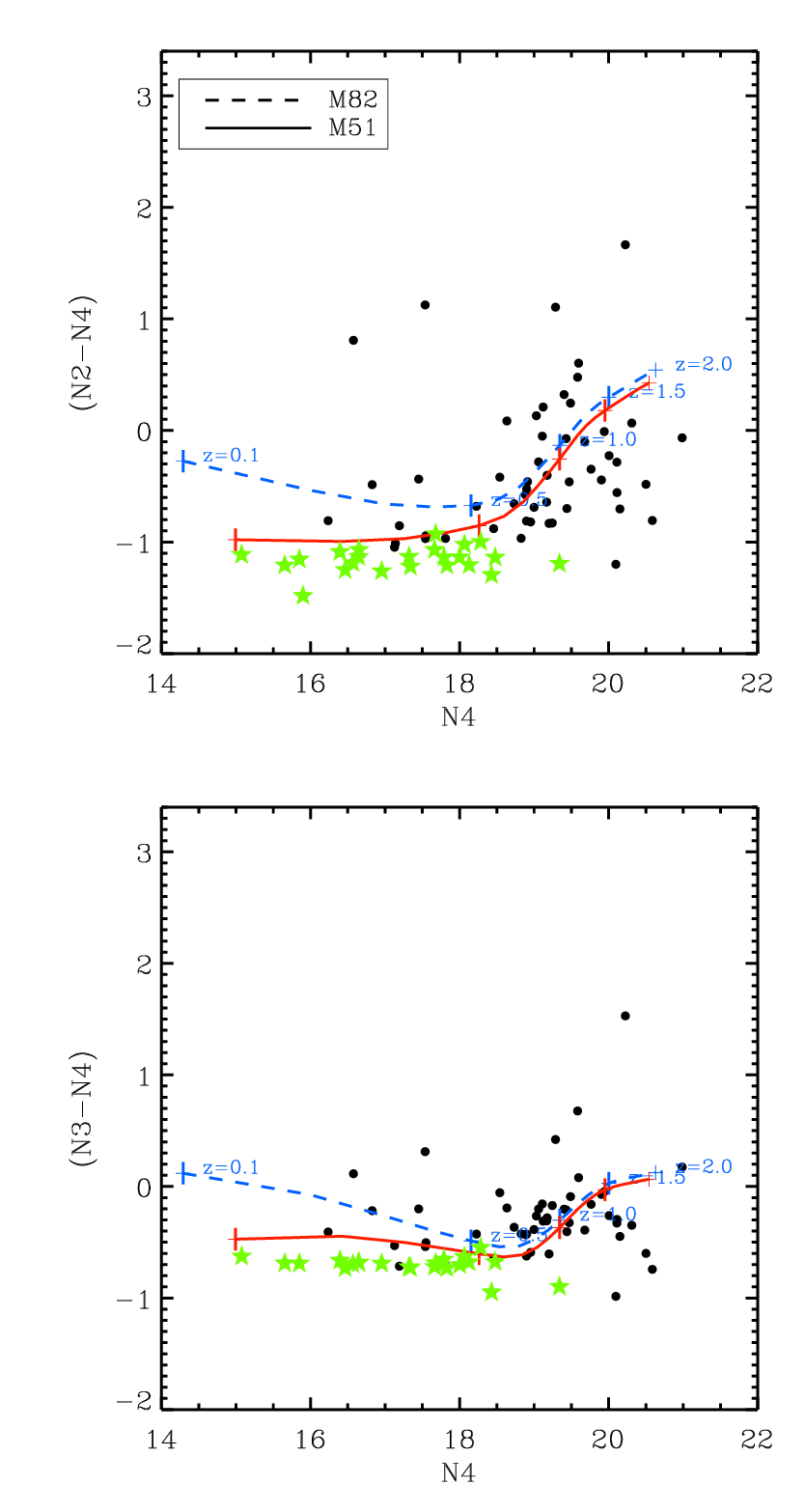}
  \end{center}
   \caption{The N2-N4 versus N4 and N3-N4 versus N4 color-magnitude diagrams.
   The sources identified as stars are marked as green stars.
 The stars lie very narrow ranges in N2-N4 and N3-N4
   colors while extragalactic objects show much broader distribution. Very red objects in
   N2-N4 color are likely to be AGNs. The locations of star forming disk galaxies at different
   redshifts are also shown as solid and dashed lines. We may conclude that the majority of
   the sources shown in these diagrams are star forming disk galaxies at low to medium redshifts.}
   \label{cm_diagram}
\end{figure}

\subsection{Characteristics of the Magnitude-limited Sample}

 So far we have been investigating the general properties of the
11 $\mu$m selected sources. We now take a closer look at
individual sources. Based on the SEDs, we divide our sources into
three categories: stars, galaxies, and AGNs.

\subsubsection{Galaxies}

  We identified galaxies from their spectral shapes,
 and their morphological appearances in the CFHT image.
 The galaxies are further divided into star-forming galaxies
 and early type galaxies based on their SED shapes.
 The total number of objects identified as star forming galaxies is 50,
 corresponding to about 70\% of the total S11 selected sample, while only five
 objects are classified as early-types.
  Fig. 6 shows SEDs of a representative set of star-forming
 galaxies at $S11 < 18.5$ mag, as well as postage stamp images
 of these galaxies.
  The SEDs are matched with 100 SED templates IR luminous galaxies in
 Chary \& Elbaz (2001), using the Bayesian photometric redshift code
 of Benitez (2000).  By doing so, we obtain an estimate of photometric
 redshift as well as the best-fit SED template. Infrared luminosity
 is then calculated by integrating the fluxes beyond the rest-frame 5 $\mu$m
 of the best-fit SED template,
 and rescaling the value according to the luminosity distance.
 The best-fit SED template is plotted over the observed points in Fig. 6,
 and we also indicate the photometric redshift in each panel.
 The probability distribution of photometric redshift is also given
 in the panel next to the SED plot. When performing the SED-fit,
 note that we have added photometric uncertainties of
 10\% for NIR, 15\% for S7 bands, and 20\% for S9W and S11 bands in order to
 account for the uncertainties in different bands that we have found
 when fitting stellar spectra (see Section 4.3.3).
  The star forming galaxies in general have SEDs that
 hump at the rest-frame 1.6 $\mu$m, and also have
 strong MIR emission in excess of the stellar light
 expected from the Rayleigh-Jeans tail of the black body radiation.
  Fig. 6 demonstrates that the S9W and S11 data points are crucial for
 sampling the redshifted excess MIR emission and the
 rest-frame 7.7 $\mu$m and 6.2 $\mu$m PAH features
 for those located at $z \gtrsim 0.2$.
  We expect that the majority of S11 or S9W selected sources in shallow
 surveys of AKARI would be star forming galaxies of this kind.
  Fig. 7 shows the SEDs of all of the objects classified as star-forming
galaxies which are brighter than 17.5 magnitude at S11,
except for the cases where AKARI fluxes are confused with their
neighbors (multiple sources).
 Like the example SEDs given in Fig. 6,  the
SEDs of 11 $\mu$m selected galaxies can be characterized by a broad peak in
near infrared and brightening at mid infrared. The mid
infrared flux can be attributed mostly to the warm dust with PAH emission
features. The SED fitting to these objects shows that they have
IR luminosity of order of $L_{IR} \sim 10^{10}$ -- $10^{12}~L_{\odot}$,
in the regime of the LIRGs. We also find that the majority of the objects
 have photometric redshifts of $0.2 \lesssim z \lesssim 0.7$.

\begin{figure}
\begin{center}
 \FigureFile(90mm, 150mm){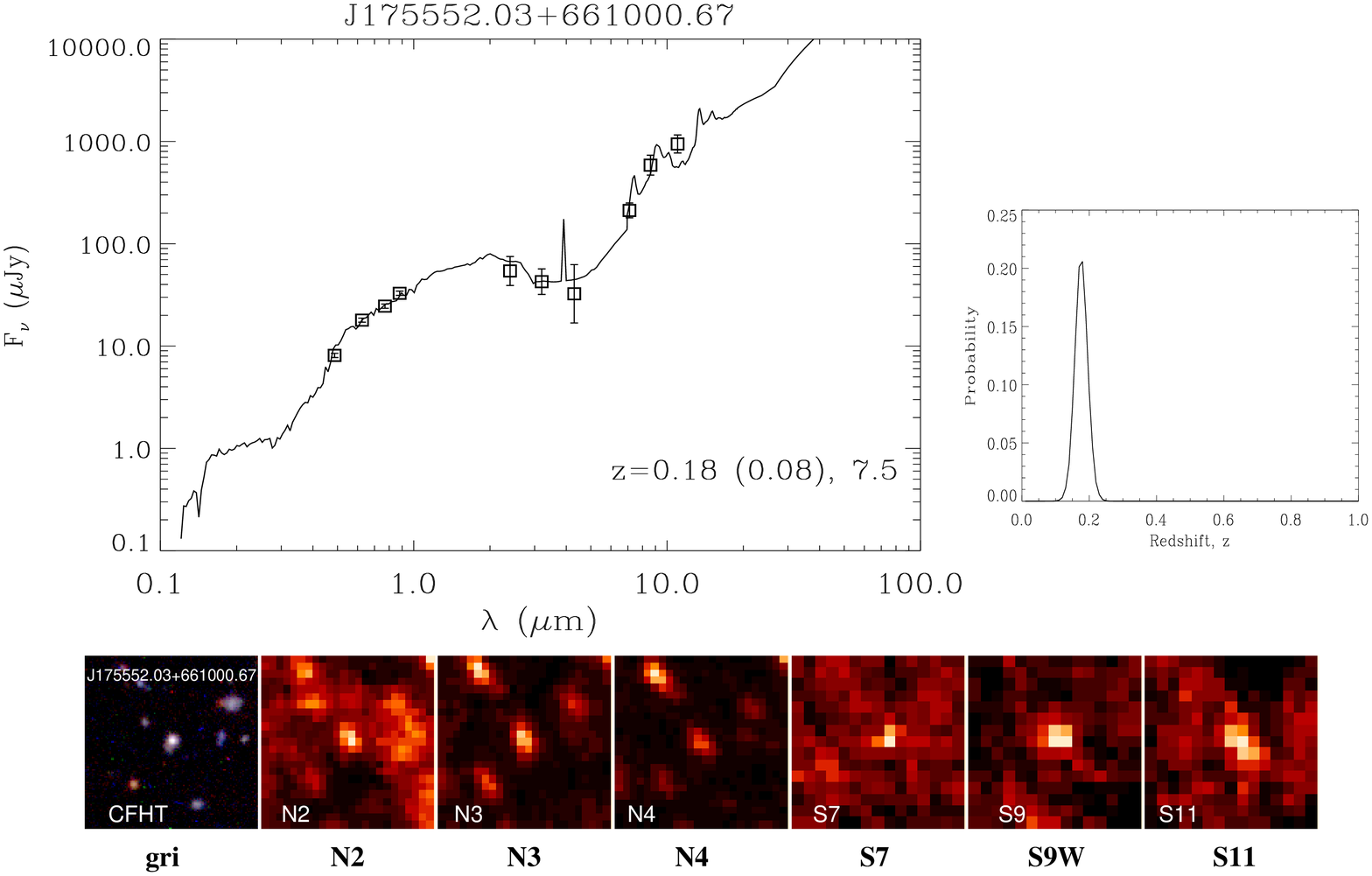}
 \FigureFile(90mm, 150mm){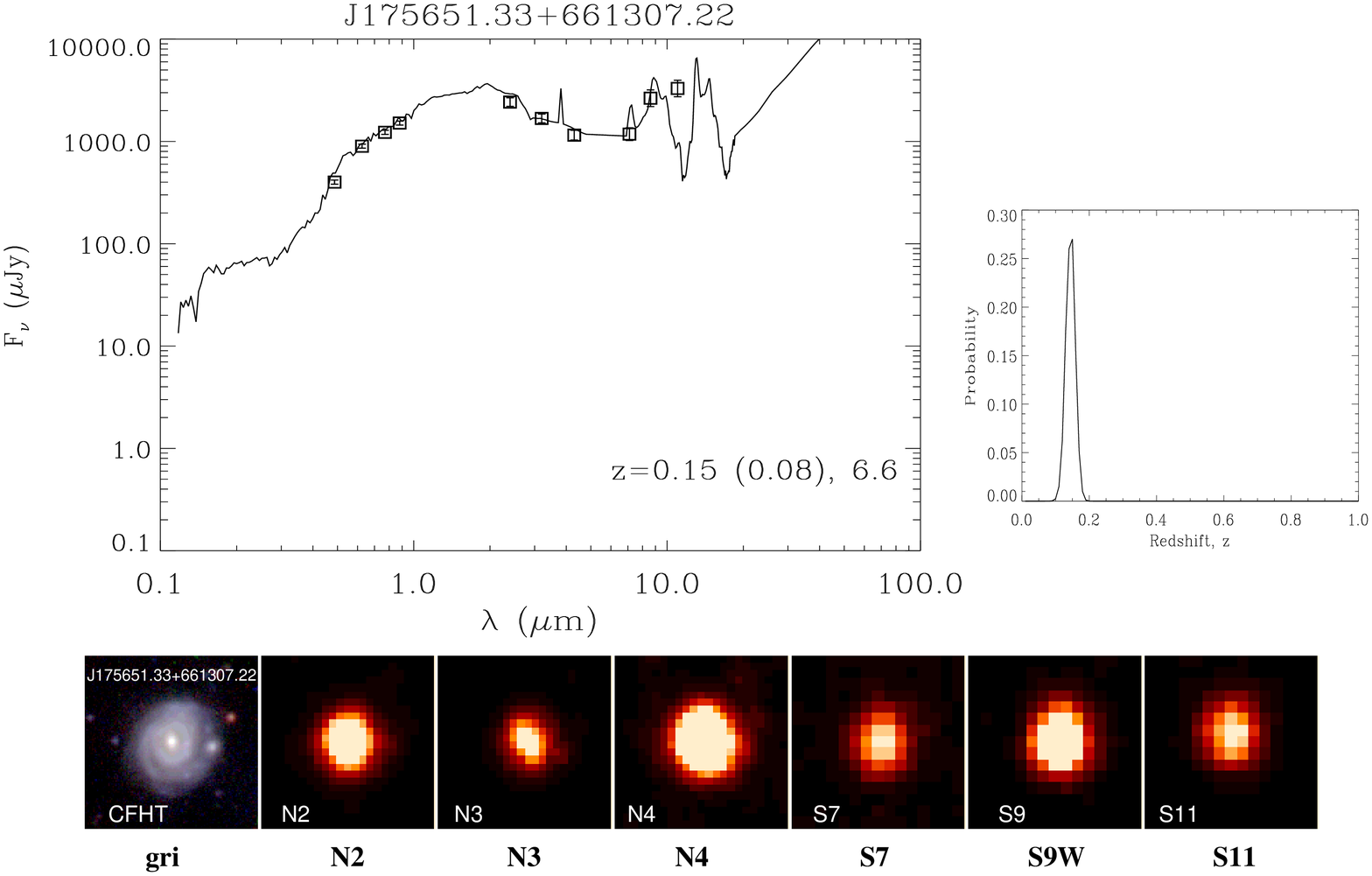}
 \FigureFile(90mm, 150mm){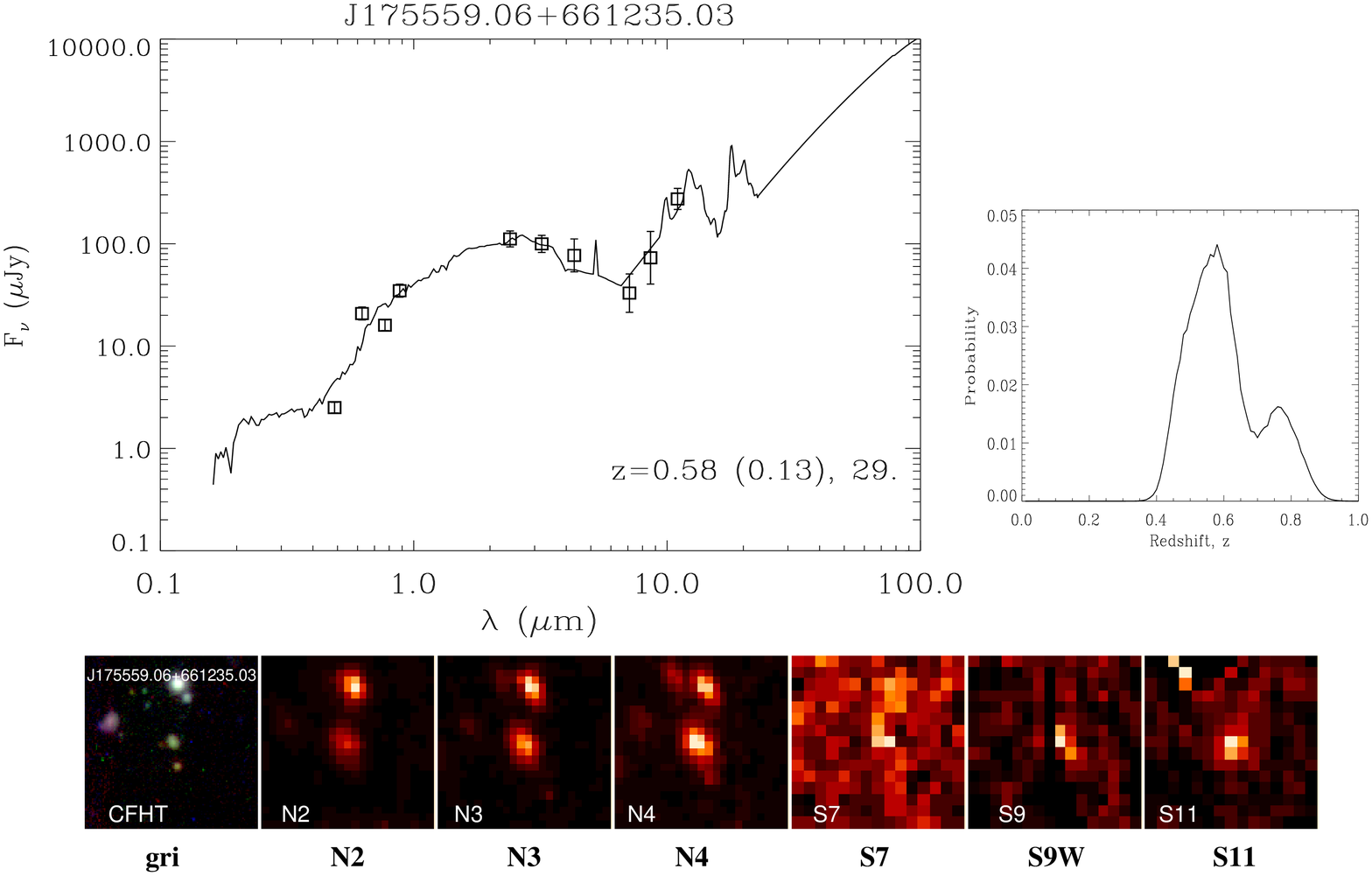}
\end{center}
\caption{The SED fitting of three star-forming galaxies: Redshfits
 are indicated in each picture, with their 1-$\sigma$ uncertainty
 shown in the parenthesis.  The number next to the
 redshift is the luminosity in units of $10^{10}$ L$_\odot$.
 For the IR luminosity,
 we estimate the uncertainty to be at least a factor of a few.
 Majority of the sources classified as star-forming galaxies lie
 in $0.2 \lesssim z \lesssim 0.7$ with luminosity ranging
 from $10^{10}$ to $10^{11}$ L$_\odot$. }
\end{figure}

\begin{figure}
 \FigureFile(150mm, 180mm){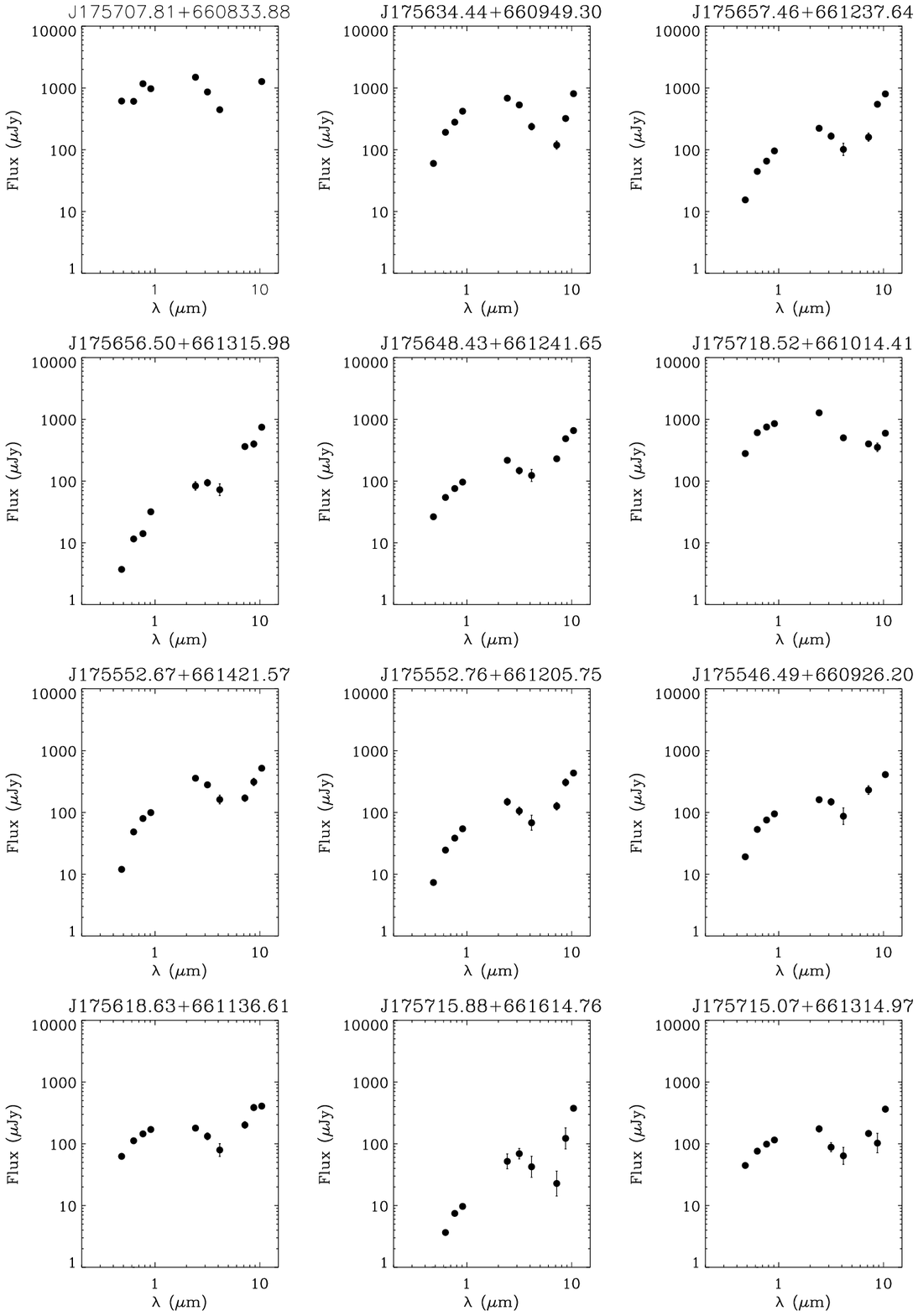}
\caption{Optical to mid infrared SEDs of 12 brightest sources that are
classified as star forming galaxies based on the inspection of CFHT image.
 We have excluded the sources whose mid-infrared fluxes might have been
contaminated by the nearby sources.
Out of 72 sources, 50 fall in this category. The SEDs are
characterized by broad peak at optical or near infrared followed
by brightening of mid infrared fluxes.}
\end{figure}

\subsubsection{AGNs}
 Among the S11 selected sample, we identify 4 objects
 ($\sim 6$ \%) with power-law SEDs which are commonly found for AGNs
 (e.g., Alonso-Herrero et al. 2006). We classify these objects
 to be AGNs. In particular, two of these objects have been
 detected in X-ray, and classified as AGNs from their X-ray
 flux as well as their optical spectra (J175609.47+661508.87
 at $z=0.6357$ and J175651.45+661242.47 at $z=1.425$;
 Gioia et al. 2003; Henry et al. 2006). The SEDs of all four ($\sim 6$\%)
sources as identified as AGN are shown in Fig. 8.
  Their IR SEDs are fitted with a power-law function
 ($f_{\nu} \sim \nu^{\alpha}$), and we find that
 the power-law index of these AGNs to be in the range
 of $-1.5 < \alpha < -0.9$, similar to that of QSOs and
 AGNs such as Mrk231 (Elvis et al. 1994;
 Ivezic et al. 2002; Alonso-Herrero et al. 2006).

\begin{figure}
\FigureFile(165mm,205mm){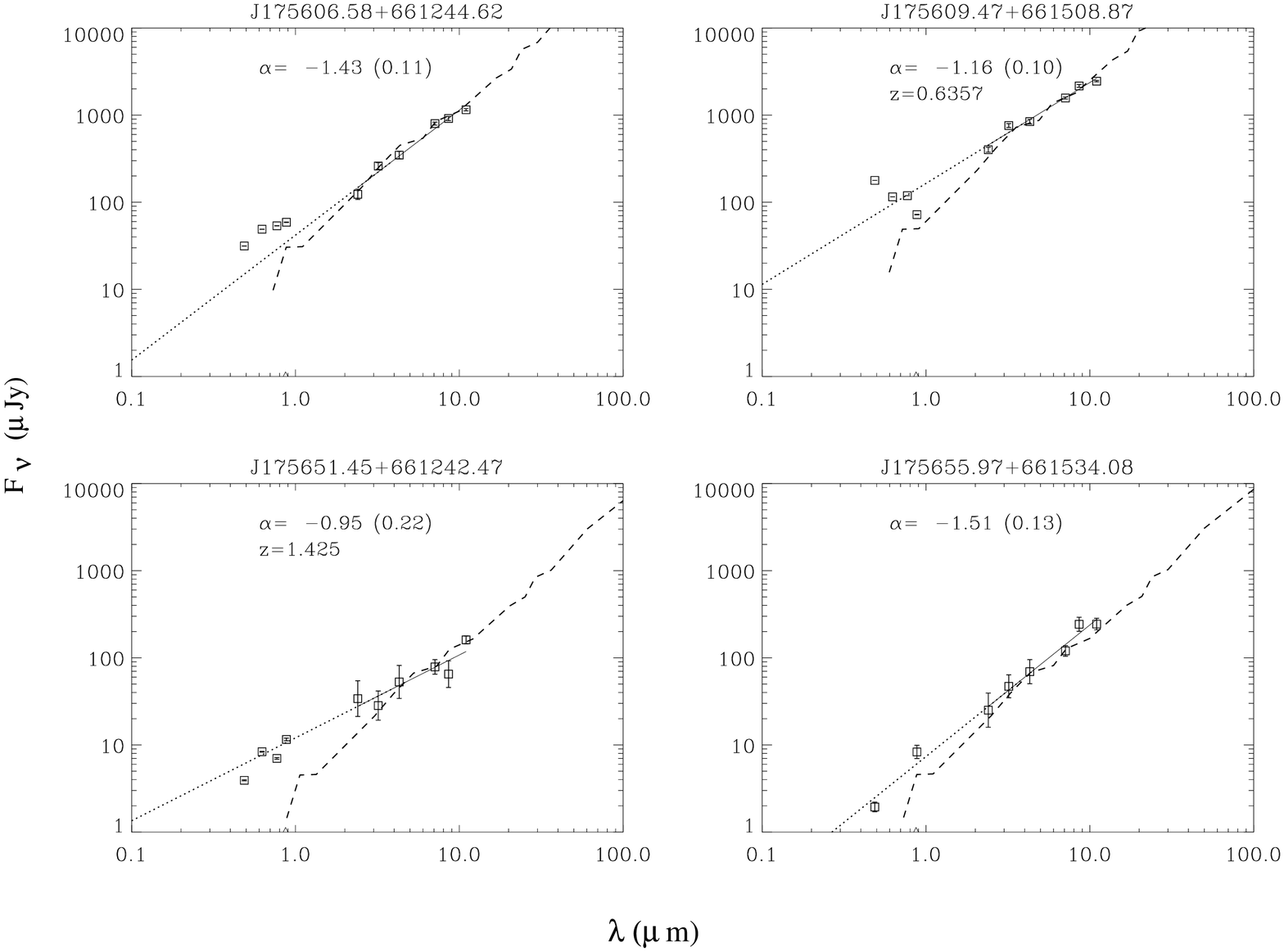}
 \caption{The SEDs of four sources which are classified as `AGNs'.
 One of them is an X-ray source located at $z=0.6357$ (J175609+661509),
 and another one is J175651.45+661242.47 at $z=1.425$.
 Also drawn in the figure are a redshifted SED of Mrk231 normalized
 at the observed 7 $\mu$m flux as  the dashed line (to the
 corresponding spectroscopic redshift, or $z=1$ if there is no
 $z_{spec}$), and the power-law fit to the observed NIR+MIR SED points
 (solid and dotted lines). The power-law index, $\alpha$ (see
 the text), is also indicated in each figure with its error shown in
 the parenthesis.
 The optical image around this source shows overdensity of faint galaxies which
 might be members of the cluster of galaxies at this redshift. }
\end{figure}

\subsubsection{Stars}

 Among 72 sources with S11 magnitude brighter than 18.5 magnitude,
14 ($\sim$ 19\%) are found to be stars. The selection of the
stellar sources are done by inspecting the cross identified CFHT
images. These are further confirmed by plotting the SEDs. The
typical stellar SEDs are shown in Fig. 9. The optical images
are often saturated, at their center, therefore, we did not
include optical fluxes in Fig. 9.
All the objects classified as stars have a steeply declining
SED shape toward the longer wavelengths, which are consistent with the Rayleigh-Jeans tail of
the black body radiation. Also plotted in Fig. 9 is a black-body fit
of the IR spectra. The figure demonstrates that the
blackbody fit is reasonable.
 The figure also demonstrates the effect of
the application of color-correction factor on photometric data points.
 The X-marks in Fig. 9 indicate the data point before the color correction.
 Without the color correction, the N4 flux is underestimated by 30\%, but the
 color correction reduces the systematic error to 10-15\% level (see also Takagi et al. 2007).
 From these fits to the stellar spectra, we estimate the systematic uncertainty in
 NIR-bands and S7-band fluxes and to be $\sim$10\%. For S9W and S11-bands,
 the uncertainty seems to be larger ($\sim$15-20\%).

\begin{figure}
\FigureFile(165mm,205mm){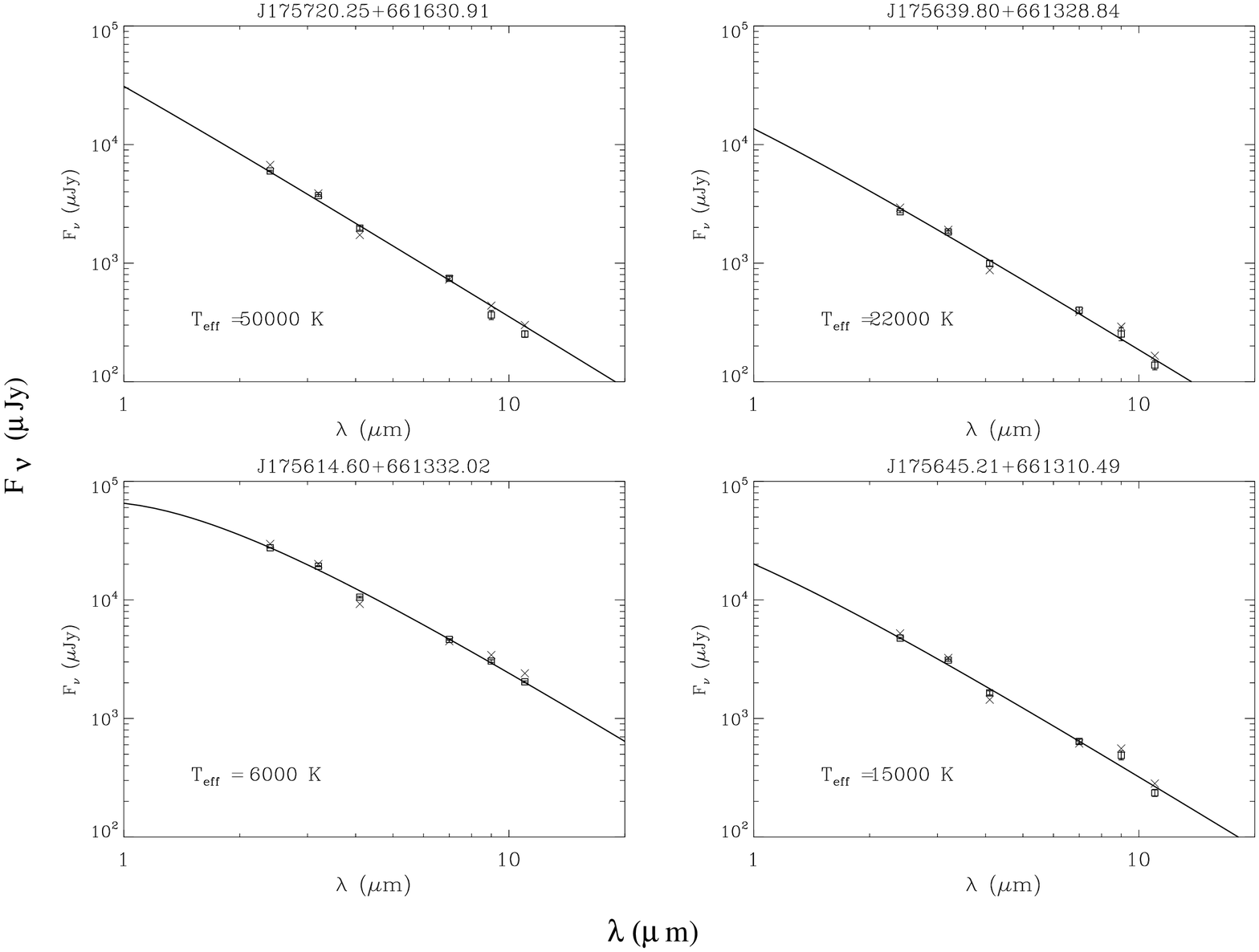}
 \caption{The SEDs of four sources which are classified as `stars' based on the
 close inspection of CFHT image (squares). Most of them are well fitted by
 the black body SED of temperatures ranging from 6000 K to
 50000 K. The size of each square represents roughly 10\% error.
 X-marks indicate the photometric data before color-correction. The color
 correction helps reduce the dip in N4-band.
}
 \end{figure}

\section{Additional 11 $\mu$m Sources}

We found several interesting sources that are beyond our
magnitude cut of 18.5 mag. Here, we discuss these objects.
SEDs and multi-band postage stamp images of these objects
are given in Fig. 10, respectively.

\begin{figure}
 \FigureFile(90mm, 105mm){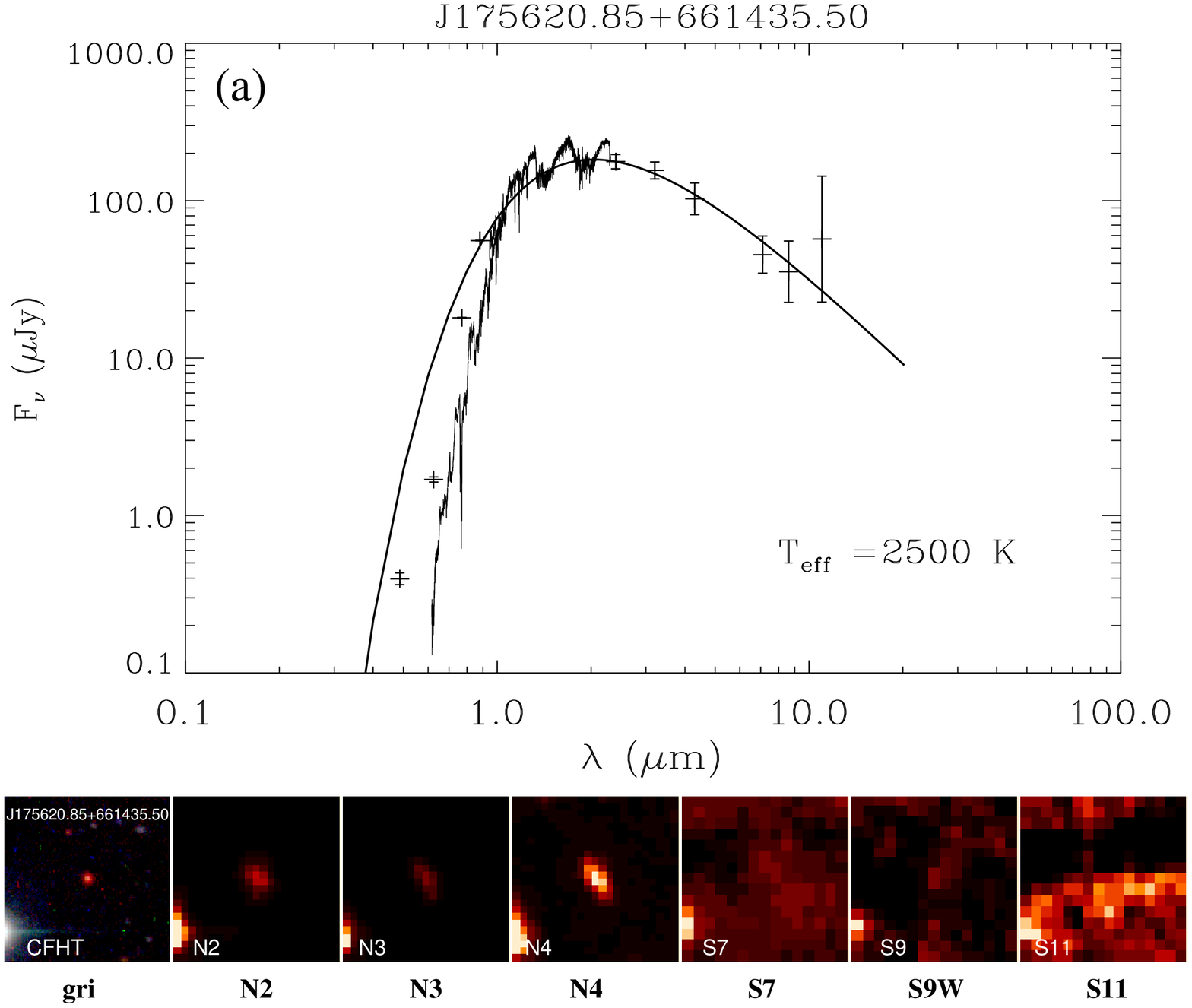}
 \FigureFile(90mm, 105mm){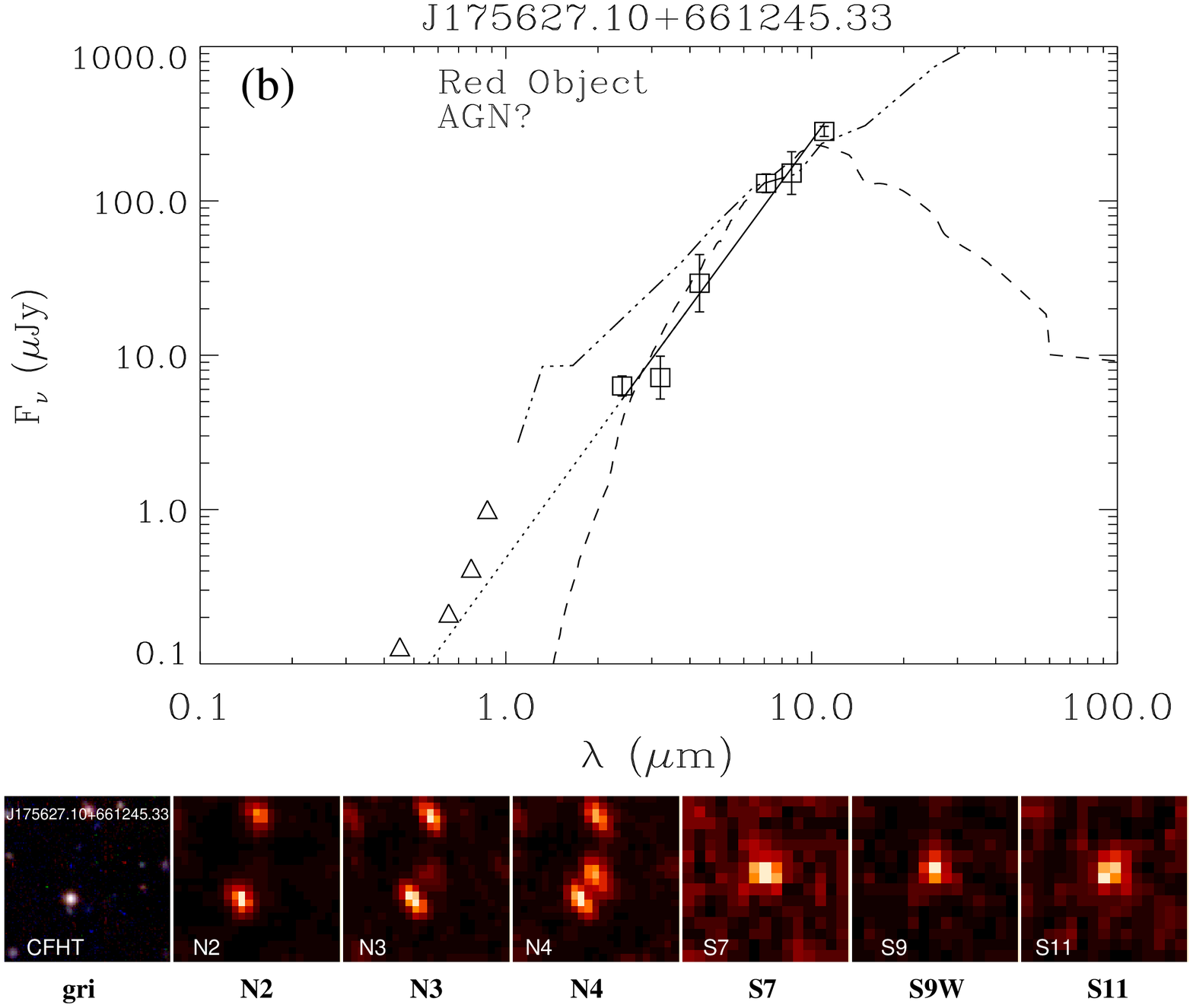}
 \FigureFile(90mm, 105mm){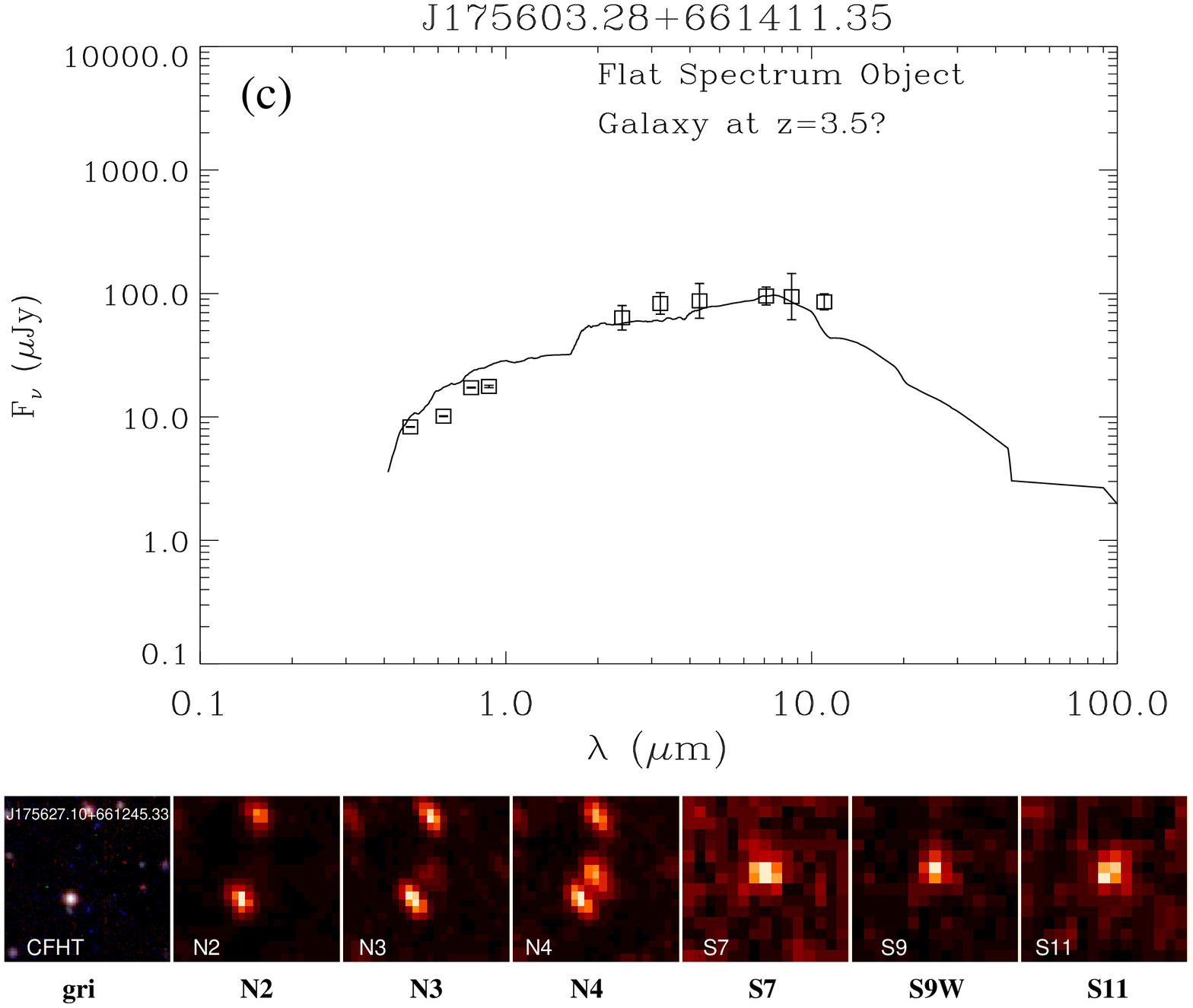}
 \FigureFile(90mm, 105mm){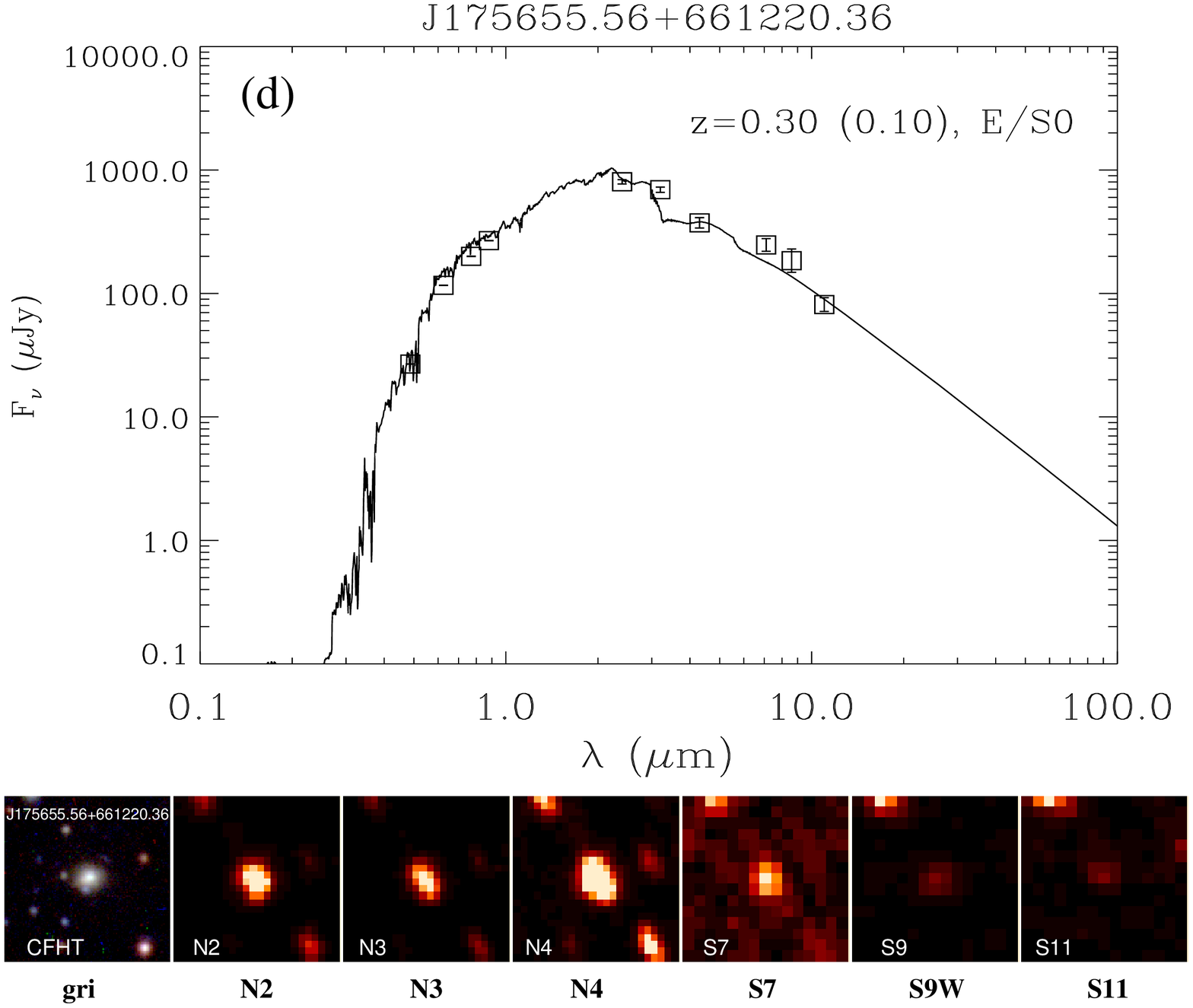}
 \caption{The spectral energy distributions
 of interesting, faint 11 $\mu$m sources. From the upper-left to
 the clockwise direction, a brown dwarf candidate (a), a dusty-AGN (b),
 a flat-spectrum object (c), and an E/S0 at $z=0.3 \pm 0.1$. (d). The
 4-$\sigma$ detection limits of CFHT image are also plotted
 as triangles. Also plotted below each figure is the postage stamp
 image of the object ($gri$-color composite of CFHT image,
 N2, N3, N4, S7, S9W, and S11 from left). The size of each postage
 stamp image is {\bf 30\farcs0 $\times$ 30\farcs}.}
 \end{figure}

\subsection{Brown Dwarf Candidate}

We find that an object J175620.85+661435.5 has a very red optical color,
and a very compact appearance in the CFHT image.
When combined with the AKARI N2, N3, N4, S7, and S9W data,
its SED can be fitted well with that of a black body with
the surface temperature of $T_{eff} \simeq 2500$ K (dashed line).
The observed SED also resembles that of a L0V-type star, 2MASS J03454316+2540233
(solid line; Kirkpatrick et al. 1999; McLean, et al. 2003).
The overall shape of the SED is very hard to fit with SEDs of
extragalactic sources, thus,
we consider J175620.85+661435.5 as a strong candidate to be a
brown dwarf.
Current limits on fluxes long-ward of 7 $\mu$m have large uncertainties.
To be able to say if this object has a circumstellar
disk like some brown dwarfs (Riaz et al. 2006; Luman et al. 2005), we
require a deeper data which we expect to accumulate at the end of
the NEP survey.

\subsection{J175627.10+661245.33: Peculiarly red object}
We find in this field, an extremely red object, J175627.10+661245.33,
which is not detected in all of the optical bands
but is visible in N2 through S11.
Fig. 10b shows the postage  stamp image of this galaxy as well as its SED.
Since they are visible in all the AKARI bands, we believe that
the detection of this object is secure.
The optical-NIR color of J175627.10+661245.33 is $(i-N2)_{AB} \gtrsim 6$,
and $(R-N2)_{AB} > 6.7$, placing it to be one of the reddest EROs and
also possibly a Hyper-Extremely Red Object (HERO; e.g., Im et al. 2002).
The most likely explanation for the nature of this source is that it is
a dust-obscured AGN. The power-low slope of this object is
$\alpha \sim -2.8$ where $\alpha$ is defined as
$f_{\nu} \sim \nu^{\alpha}$.  The power-law sources, which are dominated by
AGNs, have the power-law slopes of $\alpha = -0.5 \sim -2.8$
(Alonso-Herrero et al. 2006). This object qualifies to be in such
a category.
The spectral energy distribution of J175627.10+661245.33 can
be also fitted by that of (i) a heavily dust-extinct star forming galaxy
at $z > 5$ (dashed-line); or (ii) a few hundred Myr object at $z \sim 9.5$.
The case (ii) is unlikely, since the estimated stellar mass for
such an object exceeds above $10^{13} M_{\odot}$.

\subsection{Flat spectrum source}

We find that J175603.28+661411.35 to be a source which has
a nearly flat IR spectrum (Fig. 10c). The spectral slope of the IR points
is $\alpha \sim 0.23$, which is too flat to be a typical AGN
($-2.8 < \alpha < -0.5$; Ivezic et al. 2002; Alonso-Herrero et al. 2006).
The flat IR spectrum and the detection at $g^{'}$-band suggests that
this could be a several hundred Myr, star-forming galaxy at $z \sim 3.5$
with a moderate amount of dust extinction ($E(B-V) \sim 0.1$), although
this object could also be an AGN. The nature of this object is currently
uncertain.

\subsection{Early type galaxies}

SEDs of several objects, not in the S11 sample, are found to be consistent
with those of early-type galaxies at moderate redshift.
Fig. 10d shows one such example at the estimated distance of
$z=0.3 \pm 0.1$. Shown in the solid line is the spectrum of
a 2.5 Gyr-old passively evolving galaxy with the metallicity 2.5 times
the solar value.  No obvious excess in MIR beyond $7 \mu$m has been
found for this object. The implication is that the dust emission
from circumstellar material around AGB stars is minimal
and that this early-type galaxy is at least a few Gyrs old
from such a point of view (Piovan et al. 2003).

\section{Summary}

Using the AKARI IRC data near NEP that are obtained
for the performance evaluation purpose,
we have studied the nature of 11 $\mu$m selected sources.
The field covers roughly 10' x 10', and has a multi-wavelength
imaging data in the CFHT $g*$, $r^\prime$, $i^\prime$, and $z^\prime$,
and the AKARI N2, N3, N4, S7, S9W, and S11.
The S11 source count from the stacked images of 2-3 pointing data
shows that there are 72 objects at  $S11 < 18.5$ mag, and
that the source counts become highly incomplete beyond $S11 > 19$ mag.
We examine the nature of sources in the magnitude-limited sample
of $S11 <18.5$ mag, which we consider to be highly reliable and complete.
We find that $\sim 68$\% of the S11 sources are star-forming
galaxies at $0.2 \lesssim z \lesssim 0.7$, whose PAH emissions redshifted into
the S11 band.
The early type galaxies comprise much smaller fraction ($\sim$6\%) of
the sample.
We also classify $\sim$ 6\% of the S11 selected sources
to be AGNs. The rest are found to be stars.
 We have also examined the nature of several interesting
 sources that are fainter than  $S11 > 18.5$ mag.
 A brown dwarf candidate has been identified,
 as well as an Extremely Red Objects with no optical counterpart.
  We conclude that the AKARI S11 band is efficient for
 sampling and studying star-formation activity at moderate redshift,
 and other interesting sources such as AGNs and brown dwarfs.

\vskip 0.5cm AKARI is a JAXA project with the participation
    with ESA. The CFHT data used in this work are based on observations
    obtained with MegaPrime/MegaCam, a joint project of CFHT and CEA/DAPNIA,
    at the Canada-France-Hawaii Telescope (CFHT) which is operated by
    the National Research Council (NRC) of Canada, the Institute National
    des Sciences de l'Univers of the Centre National de
    la Recherche Scientifique of France, and the University of Hawaii.
    This work was supported in part by the KRF grant
No. R14-2002-01000-0. MI and HS were supported by KOSEF grant No.
R01-2005-000-10610-0. We thank the anonymous referee for useful
suggestions.

\end{document}